https://doi.org/10.46813/2024-151-015# CROSS-SECTIONS OF PHOTONEUTRON REACTION $^{nat}$Mo($\gamma$,xn)$^{93m}$Mo AT THE BREMSSTRAHLUNG ENERGY UP TO 95 MeV

I.S. Timchenko[1,2], O.S. Deiev[2], S.M. Olejnik[2], S.M. Potin[2], V.A. Kushnir[2],
V.V. Mytrochenko[2], S.A. Perezhogin[2]
[1]*Institute of Physics, Slovak Academy of Sciences, Bratislava, Slovakia;*
[2]*National Science Center "Kharkov Institute of Physics and Technology", Kharkiv, Ukraine*
*E-mail: timchenko@kipt.kharkov.ua; iryna.timchenko@savba.sk*The photoproduction of the $^{93m}$Mo nucleus on natural molybdenum was studied using the electron beam of the LUE-40 linac RDC "Accelerator" NSC KIPT. Measurements were performed using the activation method and off-line γ-ray spectrometric technique. For the $^{nat}$Mo($\gamma$,xn)$^{93m}$Mo reaction the experimental flux-averaged cross-section $\langle\sigma(E_{\gamma max})\rangle_m$ at the bremsstrahlung end-point energy range of 38…93 MeV was first time determined. The theoretical values of the yields $Y_{m,g}(E_{\gamma max})$ and flux-averaged cross-sections $\langle\sigma(E_{\gamma max})\rangle_{m,g}$ for the $^{nat}$Mo($\gamma$,xn)$^{93m,g}$Mo reactions were calculated using the cross-sections $\sigma(E_\gamma)$ from the TALYS1.96 code for different gamma strength functions and level density models. The comparison showed strong discrepancies between the experimental values and the theoretical results of $\langle\sigma(E_{\gamma max})\rangle_m$.

PACS: 25.20.−x, 27.60.+q## INTRODUCTION

Investigation of the mechanism of nuclear reactions (direct, compound, pre-equilibrium, etc.) is performed using experimental data on photonuclear reactions. To test and update modern computational codes, such as TALYS, EMPIRE, CCONE, COH3, used to simulate photoinduced reaction cross-sections, experimental values of the cross-sections $\sigma(E_\gamma)$ and $\langle\sigma(E_{\gamma max})\rangle$ are needed at a wide range of atomic masses and energies. However, at present, the available experimental data in the GDR region, obtained in different laboratories, are not always consistent. There is also a lack of data on photoproton and multiparticle photonuclear reactions (including emission of small clusters). This led to the emergence of works on analyzing the reliability of previously measured cross-sections and, as a consequence, stimulated new experiments.

Studies of photonuclear reactions on molybdenum isotopes have been performed in a number of papers, for example, [1–6]. Thus, in the case of the formation of the $^{93}$Mo nucleus on the $^{94}$Mo nucleus, total cross-section values measured by direct neutron detection are available in [7–9]. The work [10] shows agreement between experimental data for the $^{94}$Mo($\gamma$,n)$^{93tot}$Mo reaction and theoretical calculations performed in the TALYS1.95 code [11] with different gamma strength functions and level density models.

There are no data on the cross-section for the formation of the $^{93m}$Mo nucleus in a metastable state on natural molybdenum or other Mo stable isotopes in international databases [12]. It is possible to determine the cross-section for the photoproduction of the $^{93m}$Mo nucleus using the activation method and the off-line γ-ray spectrometric technique.

In this work, a study of the $^{93m}$Mo production in photoneutron reactions on $^{nat}$Mo at the bremsstrahlung end-point energy range of $E_{\gamma max}$ = 38…93 MeV is carried out. The obtained experimental results are compared with theoretical estimates performed with the cross-sections $\sigma(E_\gamma)$ from the TALYS1.96 code for different gamma strength functions *GSF* 1–9 and level density models *LD* 1–6.

This work continues studies of photonuclear reactions on natural molybdenum performed earlier at the linear electron accelerator LUE-40 of the Research and Development Center "Accelerator" of the National Science Center "Kharkov Institute of Physics and Technology" for the formation of $^{90}$Nb, $^{90}$Mo [13], and $^{95}$Nb [14, 15].

## 1. EXPERIMENTAL PROCEDURE

The experimental complex for the study of the formation of the $^{93m}$Mo nucleus in the photoneutron reactions on $^{nat}$Mo is presented as a block diagram in Fig. 1.

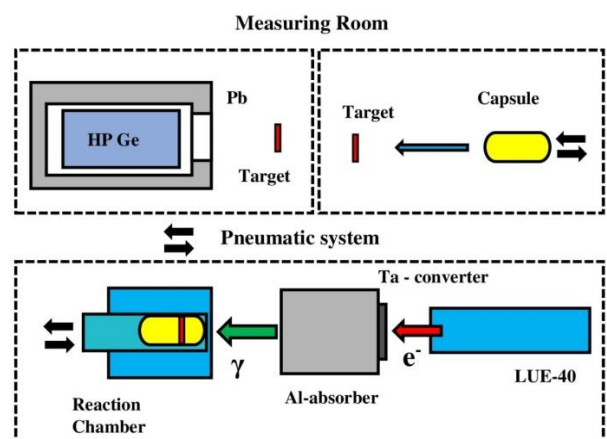

*Fig. 1. Schematic block diagram of the experiment. The upper part shows the measuring room, where the irradiated target is extracted from the capsule and is placed in front of the target the HPGe detector for induced γ-activity measurements. The lower part shows the linac LUE-40, the Ta converter, the Al absorber, and the exposure reaction chamber*

The experiment was performed using the electron linear accelerator LUE-40 of RDC "Accelerator" NSC KIPT. The linac LUE-40 provides an electron beam

ISSN 1562-6016. Problems of Atomic Science and Technology. 2024. №3(151)    15

with an average current $I_e \approx 3$ μA and full width at half maximum (FWHM) of energy spectrum $\Delta E_e/E_e \approx 1\%$. The range of initial energies of electrons $E_e = 35…95$ MeV. A detailed description and parameters of the linac are given in works [16–19].

On the axis of the electron beam, there are a converter, an absorber, and a reaction chamber. The converter, made of tantalum metal, is a 20×20 mm plate with thickness $l = 1.05$ mm, and is attached to an aluminum absorber, shaped as a cylinder, with dimensions ⌀ 100×150 mm. The thickness of the aluminum absorber was calculated to clean the beam of γ-quanta from electrons with energies up to 100 MeV.

For the experiment, targets were made of natural molybdenum, which were thin discs with a diameter of 8 mm and a thickness of ~0.11 mm, which corresponded to a mass of 57…60 mg. Natural molybdenum consists of 7 stable isotopes, with isotope abundance, %: $^{92}$Mo – 14.84, $^{94}$Mo – 9.25, $^{95}$Mo – 15.92, $^{96}$Mo – 16.68, $^{97}$Mo – 9.55, $^{98}$Mo – 24.13, $^{100}$Mo – 9.63 (according to [11, 20]).

The target was placed in an aluminum capsule and delivered by a pneumatic transport system to the reaction chamber for irradiation and back to the measuring room to record the induced γ-activity of reaction products in the target substance.

The γ-quanta of the reaction products were detected using a Canberra GC-2018 semiconductor HPGe detector. Its efficiency was 20% relative to the NaI(Tl) scintillator with dimensions 3 inches in diameter and 3 inches in thickness at energy $E_\gamma = 1332$ keV. The resolution FWHM is 1.8 keV for energy $E_\gamma = 1332$, and 0.8 keV for $E_\gamma = 122$ keV. The dead time for γ-quanta detection varied between 0.1 and 5%. The absolute detection efficiency $\varepsilon(E_\gamma)$ for γ-quanta of different energies was obtained using a standard set of γ-rays sources: $^{22}$Na, $^{60}$Co, $^{137}$Cs, $^{152}$Eu, $^{241}$Am, $^{133}$Ba. The analytical curve in the form $\ln\varepsilon(E_\gamma) = \Sigma a_i (\ln E_\gamma)^i$, proposed in [21], was used to determine the value of $\varepsilon(E_\gamma)$ for various energies of γ-quanta.

The electron bremsstrahlung spectra were calculated using the open-source software code GEANT4.9.2, PhysList G4LowEnergy [20]. The real geometry of the experiment was used in calculations as well as the space and energy distributions of the electron beam were taken into account.

The bremsstrahlung flux was monitored by the yield of the $^{100}$Mo(γ,n)$^{99}$Mo reaction (the half-life $T_{1/2}$ of the $^{99}$Mo nucleus is (65.94 ± 0.01) h) by comparing the experimentally obtained flux-averaged cross-section $\langle\sigma(E_{\gamma max})\rangle$ with the computation data $\langle\sigma(E_{\gamma max})\rangle_{th}$. To determine the experimental $\langle\sigma(E_{\gamma max})\rangle$ values it has used the yield for the γ-line of energy $E_\gamma = 739.50$ keV and intensity $I_\gamma = (12.13 ± 0.12)\%$. The flux-averaged cross-section $\langle\sigma(E_{\gamma max})\rangle_{th}$ values were computed with the cross-sections $\sigma(E_\gamma)$ from the TALYS1.96 code. Details of the monitoring procedure can be found in [22–26].

To study the $^{nat}$Mo(γ,xn)$^{93m}$Mo reaction (the half-life $T_{1/2}$ of the $^{93m}$Mo nucleus is (6.85 ± 0.07) h) the yield for the γ-line of energy $E_\gamma = 684.67$ keV and intensity $I_\gamma = (99.7 ± 0.2)\%$ was used. Nuclear spectroscopic data of the nuclei-products of reactions were taken from the database [27].

The metastable level has a high excitation energy and spin, the values of which are 2.425 MeV and 21/2+, respectively. Such parameters probably will lead to low values of the reaction yield and, accordingly, the low cross-section of the $^{nat}$Mo(γ,xn)$^{93m}$Mo reaction.

The γ-radiation spectrum of a $^{nat}$Mo target irradiated by a beam of bremsstrahlung γ-quanta with high endpoint energy is a complex pattern. There are emission γ-lines of nuclei-product of the $^{nat}$Mo(γ,ypxn) reactions located on a background substrate, which is formed as a result of Compton scattering of photons. As an example, in Fig. 2 γ-radiation spectrum of a $^{nat}$Mo target with a mass of 57.862 mg after irradiation with $E_{\gamma max} = 92.50$ MeV is shown.

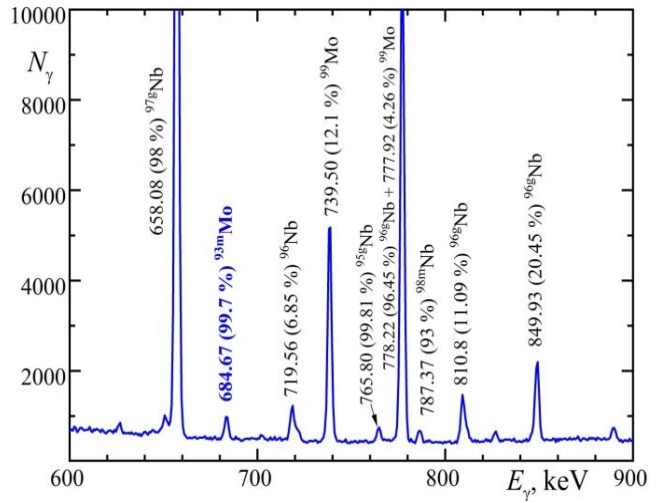

*Fig. 2. Fragment of γ-ray spectrum in the energy range $600 \leq E_\gamma \leq 900$ keV from the $^{nat}$Mo target of mass 57.862 mg after irradiation of the bremsstrahlung γ-flux at $E_{\gamma max} = 92.50$ MeV. Irradiation and measurement time were both 3600 s. The γ-line marked by blue color was used in the investigation*

## 2. CALCULATION OF CROSS-SECTIONS $\sigma(E_\gamma)$, YIELDS $Y_i(E_{\gamma max})$ AND FLUX-AVERAGED CROSS-SECTIONS $\langle\sigma(E_{\gamma max})\rangle$

The $^{93m}$Mo nucleus in the metastable states can be formed in photoneutron reactions on $^{nat}$Mo. Natural molybdenum consists of 7 stable isotopes, but only six isotopes contribute to the formation of the $^{93m}$Mo nucleus, respectively, there are six reactions with thresholds:
$^{94}$Mo(γ,n)$^{93m}$Mo – $E_{thr} = 12.10$ MeV;
$^{95}$Mo(γ,2n)$^{93m}$Mo – $E_{thr} = 19.47$ MeV;
$^{96}$Mo(γ,3n)$^{93m}$Mo – $E_{thr} = 28.63$ MeV;
$^{97}$Mo(γ,4n)$^{93m}$Mo – $E_{thr} = 35.45$ MeV;
$^{98}$Mo(γ,5n)$^{93m}$Mo – $E_{thr} = 44.09$ MeV;
$^{100}$Mo(γ,7n)$^{93m}$Mo – $E_{thr} = 58.31$ MeV.

The thresholds for the formation of the $^{93m}$Mo nucleus in the metastable state are higher than in the ground state with an excess excitation energy of 2.425 MeV.

The cross-sections $\sigma(E_\gamma)$ of studied reactions for monochromatic photons were calculated using the TALYS1.96 code [11] for the six different level density models *LD* and nine gamma strength functions *GSF*.

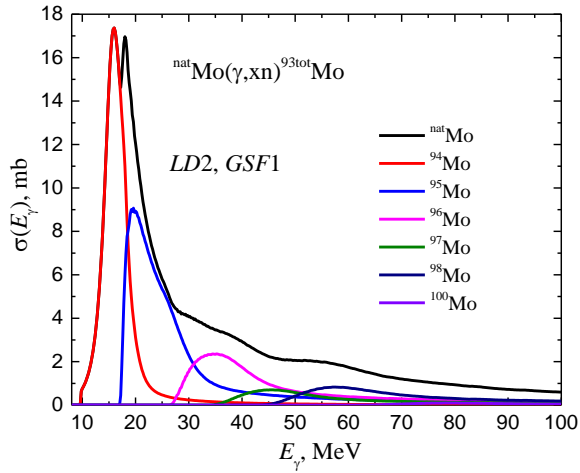

*Fig. 3. Theoretical cross-sections σ($E_\gamma$) for the formation of the $^{93tot}$Mo nucleus on 6 stable isotopes of molybdenum (taking into account isotope abundance) and on $^{nat}$Mo (black curve). The calculations were performed in the TALYS1.96 code for options LD2, GSF1*

There are three phenomenological level density models and three options for microscopic level densities:
*LD*1: Constant temperature + Fermi gas model, introduced by Gilbert and Cameron.
*LD*2: Back-shifted Fermi gas model.
*LD*3: Generalized superfluid model.
*LD*4: Microscopic level densities (Skyrme force) from Goriely's tables.
*LD*5: Microscopic level densities (Skyrme force) from Hilaire's combinatorial tables.
*LD*6: Microscopic level densities based on temperature dependent Hartree-Fock-Bogoliubov calculations using the Gogny force from Hilaire's combinatorial tables.

The shape of the excitation function curve mainly depends on the gamma-strength functions *GSF*. There are nine gamma strength functions available in the TALYS 1.96 code namely:
*GSF*1: Generalized Lorentzian of Kopecky and Uhl.
*GSF*2: Generalized Lorentzian of Brink and Axel.
*GSF*3: Skyrme-Hartree-Fock + BCS approximation.
*GSF*4: Hartree-Fock-Bogolyubov-quasiparticle random-phase approximation (HFB).
*GSF*5: Hybrid model of Goriely.
*GSF*6: Goriely *T*-dependent HFB.
*GSF*7: *T*-dependent Relativistic Mean Field model.
*GSF*8: Gogny D1M HFB + QRPA.
*GSF*9: Simplified Modified Lorentzian.

The calculated cross-sections σ($E_\gamma$) for the formation of the $^{93tot}$Mo nucleus on 6 stable isotopes of molybdenum (94, 95, 96, 97, 98, 100) are shown in Fig. 3. The cross-sections are given taking into account the abundance of isotopes. The cross-sections for natural molybdenum (black curve) were obtained as the sum of the cross-sections for 6 isotopes with an abundance of isotopes. Two dominant contributions to the total cross-section of the $^{nat}$Mo(γ,xn)$^{93tot}$Mo reaction are observed from isotopes 94 and 95.

The contribution of isotopes changes noticeably when calculating the cross-section for the formation of $^{93m}$Mo in the metastable state. The calculation results for various model options are shown in Fig. 4.

Using the theoretical cross-section σ($E_\gamma$), one can obtain the reaction yield, which is determined by the formula:

$$Y(E_{\gamma\max}) = N_n \int_{E_{thr}}^{E_{\gamma\max}} \sigma(E_\gamma) W(E_\gamma, E_{\gamma\max}) dE_\gamma, \quad (1)$$

where $N_n$ is the number of atoms of the element under study; $W(E_\gamma, E_{\gamma\max})$ is the bremsstrahlung γ-flux; $E_{thr}$ – the energy of the reaction threshold, and $E_{\gamma\max}$ – the bremsstrahlung end-point energy.

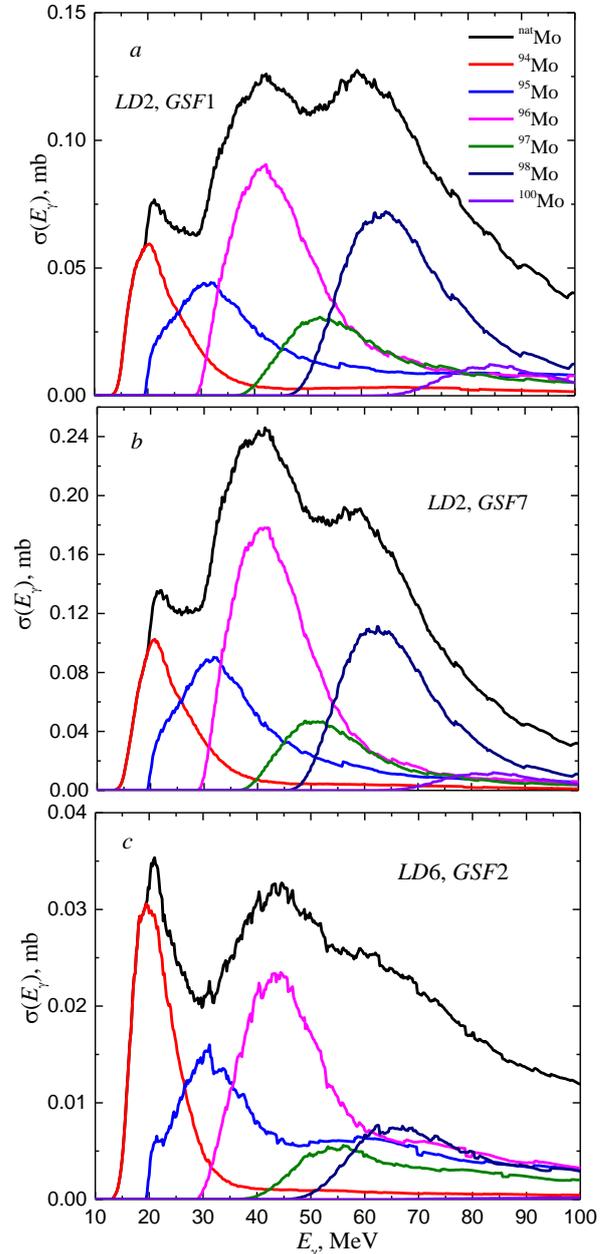

*Fig. 4. Theoretical cross-sections σ($E_\gamma$) for the formation of the $^{93m}$Mo nucleus on 6 stable isotopes of molybdenum (taking into account isotope abundance) and on $^{nat}$Mo (black curve). The calculations were performed in the TALYS1.96 code for options: LD2, GSF1 (a); LD2, GSF7 (b); LD6, GSF2 (c)*

For the estimation of the contribution of a reaction in the total production of a studied nuclide (for example, the $^{94}$Mo(γ,n) reaction in the production of the $^{93m}$Mo nucleus on $^{nat}$Mo), the normalized reaction yield

$Y_i(E_{\gamma max})$ was used. For calculation of $Y_i(E_{\gamma max})$, it was used the expression:

$$Y_i(E_{\gamma max}) = \frac{A_i \int_{E_{thr}^i}^{E_{\gamma max}} \sigma_i(E_\gamma) W(E_\gamma, E_{\gamma max}) dE_\gamma}{\sum_{k=1}^{6} A_k \int_{E_{thr}^k}^{E_{\gamma max}} \sigma_k(E_\gamma) W(E_\gamma, E_{\gamma max}) dE_\gamma}, \quad (2)$$

where $\sigma_k(E_\gamma)$ is the cross-section for the formation of the $^{93m}$Mo nucleus on the $k$-th isotope with isotopic abundance $A_k$. Summation over $k$ goes over 6 stable molybdenum isotopes $^{94,95,96,97,98,100}$Mo.

As a rule, in the presence of several isotopes, there is one whose contribution to the reaction yield dominates (> 90%), as shown, for example, in [25, 28].

In the case of the $^{nat}$Mo($\gamma$,xn)$^{93m}$Mo reaction at energies up to 20 MeV, the contribution of $^{94}$Mo is 100% (Table 1). With increasing energy, the contribution from isotope $^{94}$Mo decreases, and at 95 MeV it reaches 24 to 38%, depending on the model options. So, in the case of the formation of the $^{93m}$Mo nucleus at bremsstrahlung end-point energy above 30 MeV, it is difficult to determine the dominant reaction. For the estimation of the production of the $^{93}$Mo nucleus (metastable or ground states) in the $^{nat}$Mo($\gamma$,xn) reaction, it is necessary to take into account the contributions of six stable isotopes.

Fig. 5 shows the calculated cross-sections $\sigma(E_\gamma)$ for the formation of the $^{93g,m}$Mo nucleus. The difference between the cross-sections in the ground and metastable states reaches two orders of magnitude. In addition, the cross-section for the reaction with the formation of a metastable state strongly depends on the model parameters.

The cross-sections $\sigma(E_\gamma)$ for the $^{94}$Mo($\gamma$,n)$^{93tot}$Mo reaction, calculated in TALYS1.96 are shown in Fig. 6. As can be seen from the figure, when using one model of the strength function $GSF$1, calculations for different level density models $LD$ 1–6 give similar values. At the same time, the use of different $GSF$ 1–8 shows a scatter of $\sigma(E_\gamma)$ values. The differences in the maximum GDR were up to 20%. This allows us to select the calculation option that is the closest to experimental data. In [10], experimental data [7] were compared with calculations in the TALYS1.95 code and it was shown that the best agreement was achieved for the case of $LD$2, $GSF$1.

The situation changes when calculating the cross-section for the production of $^{93m}$Mo in a metastable state. Calculation with changes in the $LD$ and $GSF$ parameters leads to a significant scattering of the cross-sections $\sigma(E_\gamma)$. Fig. 7 shows theoretical cross-sections for the $^{94}$Mo($\gamma$,n)$^{93m}$Mo reaction at the GDR energy. The cross-section lies in the range of 0.3…1.2 mb, i.e. it differs by 4 times.

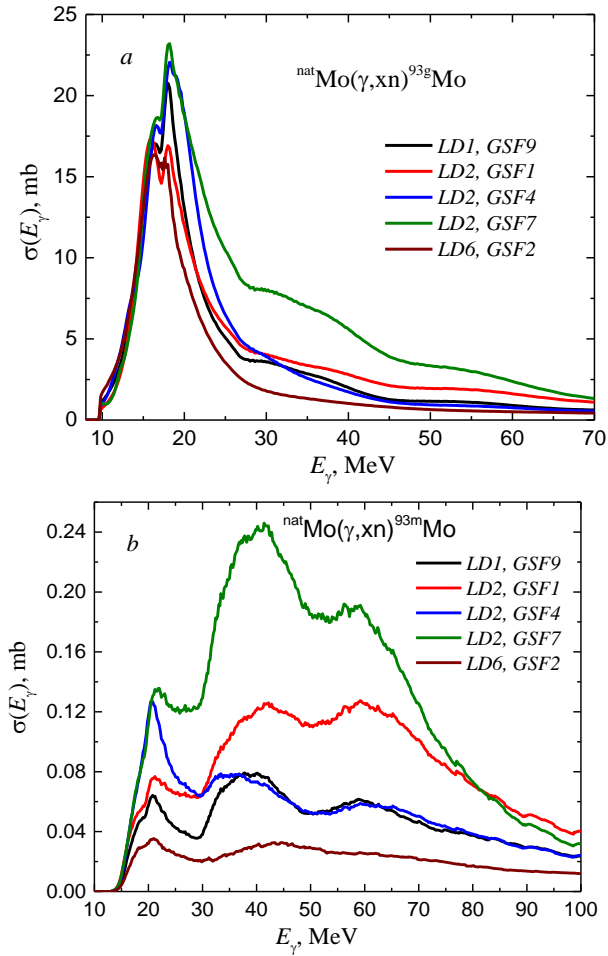

*Fig. 5. Theoretical cross-sections $\sigma(E_\gamma)$ for the formation of the $^{93g}$Mo (a) and $^{93m}$Mo (b) nuclei on $^{nat}$Mo*

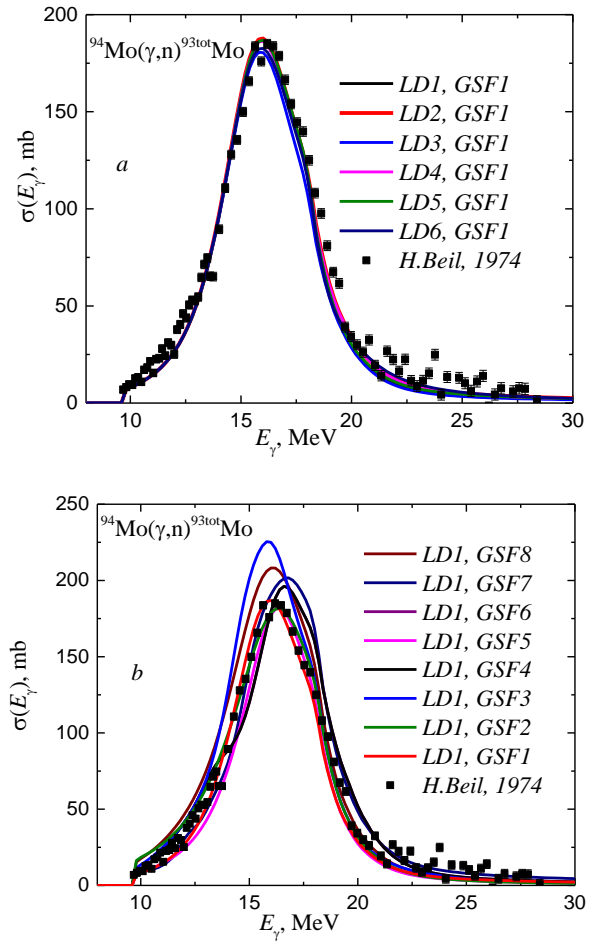

*Fig. 6. The cross-sections $\sigma(E_\gamma)$ for the $^{94}$Mo($\gamma$,n)$^{93tot}$Mo reaction, calculated in the TALYS1.96 code for options: a – LD 1–6, GSF1; b – LD1, GSF 1–8. Squares – experimental data from [7]*

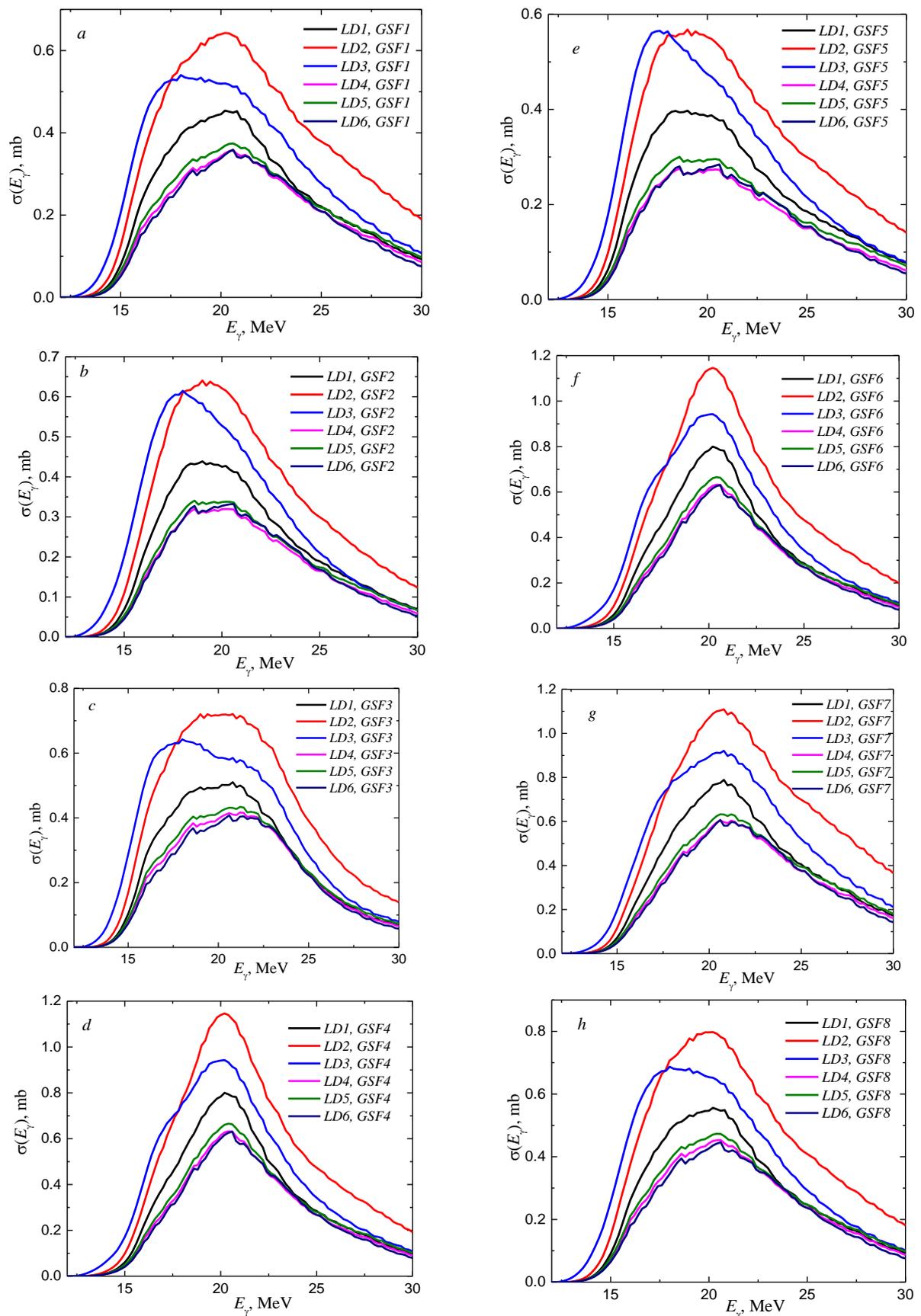

*Fig. 7. Theoretical cross-sections σ(E_γ) for the $^{93m}$Mo formation in the $^{94}$Mo(γ,n)$^{93m}$Mo reaction*

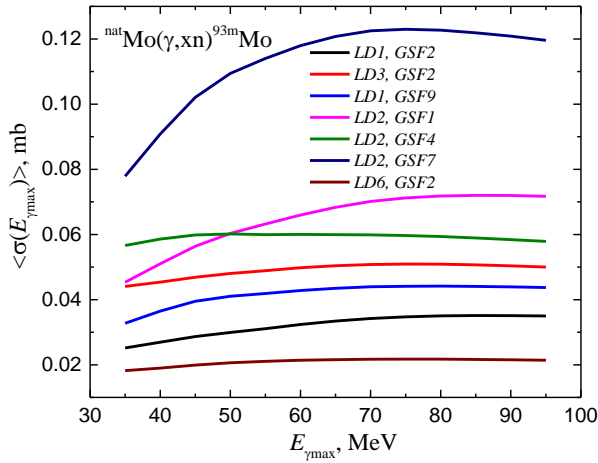

*Fig. 8. Theoretical flux-averaged cross-section ⟨σ(E$_{γmax}$)⟩$_m$ for the formation of the $^{93m}$Mo nucleus on $^{nat}$Mo*

The cross-sections σ(E$_γ$) are averaged over the bremsstrahlung flux $W(E_γ, E_{γmax})$ in the energy range from the threshold of the corresponding reaction $E_{thr}$ to the maximum energy of the bremsstrahlung spectrum $E_{γmax}$. As a result, flux-averaged cross-section values were obtained:

$$\langle \sigma(E_{\gamma\max}) \rangle = \frac{\int_{E_{th}}^{E_{\gamma\max}} \sigma(E_\gamma) W(E_\gamma, E_{\gamma\max}) dE_\gamma}{\int_{E_{th}}^{E_{\gamma\max}} W(E_\gamma, E_{\gamma\max}) dE_\gamma}. \quad (3)$$

For calculation of flux-averaged cross-sections ⟨σ(E$_{γmax}$)⟩ for the $^{nat}$Mo(γ,xn)$^{93m}$Mo reactions, we average the cross-sections σ(E$_γ$) using the minimal threshold of the studied reaction − $E_{thr}$ = 12.10 MeV. This was done to be able to compare the theoretical flux-averaged cross-sections with experimental results, where the total flux from the $^{94}$Mo(γ,n) reaction threshold to $E_{γmax}$ is used.

Fig. 8 shows cross-section ⟨σ(E$_{γmax}$)⟩$_m$ for the formation of the $^{93m}$Mo nucleus on $^{nat}$Mo for different models of *LD* and *GSF*.

## 3. EXPERIMENTAL RESULTS

The cross-sections for the formation of the $^{93m}$Mo nucleus in the metastable state in the reaction on $^{nat}$Mo can be determined from direct measurements of the number of counts of γ-quanta ΔA in the full absorption peak at an energy of 684.67 keV. To calculate the experimental values ⟨σ(E$_{γmax}$)⟩ the following expression was used:

$$\langle \sigma(E_{\gamma\max}) \rangle = \frac{\lambda \Delta A}{\varepsilon N_x I_\gamma \Phi(E_{\gamma\max})(1-e^{-\lambda t_{irr}})e^{-\lambda t_{cool}}(1-e^{-\lambda t_{meas}})}, \quad (4)$$

where $\Phi(E_{\gamma\max}) = \int_{E_{thr}}^{E_{\gamma\max}} W(E_\gamma, E_{\gamma\max}) dE_\gamma$ – the sum of bremsstrahlung flux quanta in the energy range from the reaction threshold $E_{thr}$ = 12.10 MeV to $E_{γmax}$; $N_x$ is the number of studied atoms (including 6 isotopes – 94–98, and 100); $I_γ$ – the intensity of the analyzed γ-quanta; ε – the absolute detection efficiency for the analyzed γ-quanta energy; λ is the decay constant (ln2/$T_{1/2}$); $t_{irr}$, $t_{cool}$, and $t_{meas}$ are the irradiation time, cooling time and measurement time, respectively.

The experimental values of the flux-averaged cross-section ⟨σ(E$_{γmax}$)⟩$_m$ of the $^{nat}$Mo(γ,xn)$^{93m}$Mo reaction were determined at the bremsstrahlung end-point energy of 38…93 MeV (Fig. 9 and Table 2).

The uncertainty of measured flux-averaged cross-sections was determined as a square root of the quadratic sum of statistical and systematic errors. The statistical error in the observed γ-activity is mainly due to statistics in the full absorption peak of the corresponding γ-ray, which varies between 4.9 to 13.3%. The measured ΔA value of the investigated γ ray depends on the detection efficiency, half-life, and the intensity $I_γ$. The background is generally governed by the contribution from the Compton scattering of the emitted γ-rays.

The systematical errors are due to the following uncertainties of:

1 – exposure time and the electron current ∼0.5%;
2 – γ-ray detection efficiency of the detector 2...3%;
3 – the half-life $T_{1/2}$ of the reaction products and the intensity $I_γ$ of the analyzed γ-rays;
4 – normalization of the experimental data to the yield of the monitoring reaction $^{100}$Mo(γ,n)$^{99}$Mo made up 6%. It should be noted that the systematic error in yield monitoring of the $^{100}$Mo(γ,n)$^{99}$Mo reaction stems from three errors, each reaching up to 1%. These are the statistical error in the determination of the number of counts under the γ-ray peak used for normalization, the uncertainty in the isotopic composition of natural molybdenum and in the intensity $I_γ$ used. In our calculations, we have used the percentage value of $^{100}$Mo isotope abundance equal to 9.63% [20].

The total uncertainties of the measured flux-averaged cross-sections are given in Fig. 9 and Table 2.

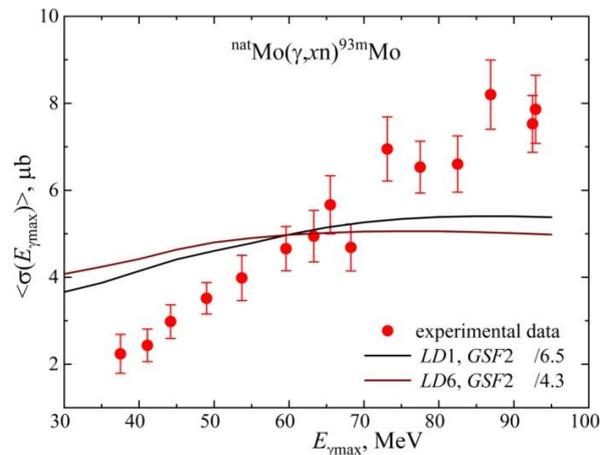

*Fig. 9. The flux-averaged cross-section ⟨σ(E$_{γmax}$)⟩$_m$ for the formation of the $^{93m}$Mo nucleus on $^{nat}$Mo. Black line – calculation for options LD1, GSF2 divided by 6.5, brown line – calculation LD6, GSF2 divided by 4.3. Red points – experimental data*

The measured value of the flux-averaged cross-section is several μb. The probable reason for such a small cross-section is the high excitation energy of the metastable state and its high spin. Such experiments can be carried out thanks to the high density of bremsstrahlung flux. Previously, such small cross-sections were measured in [29] for a photoproton reaction with the formation of a metastable state of the hafnium nucleus (30…40 μb) and in [30] for a multiparticle photonuclear reaction on aluminum with the emission of particle clusters (20…200 μb).

Fig. 9 shows also the calculated values of the flux-averaged cross-sections. There is a strong discrepancy between the theoretical cross-sections and the experimental data for all used options of the TALYS1.96 code. The difference between the slopes of the calculated curves and the experimental cross-sections may be because the ratio of the theoretical contributions of the photoproduction of the $^{93m}$Mo nucleus on stable Mo isotopes does not correspond to the observed one.

Table 1

The $^{94}$Mo(γ,n) normalized reaction yield of the formation of the $^{93m}$Mo nucleus on $^{nat}$Mo

| $E_{\gamma max}$, MeV | LD2, GSF1 | LD2, GSF7 | LD6, GSF2 |
|---|---|---|---|
| 20 | 1.00 | 1.00 | 1.00 |
| 25 | 0.86 | 0.84 | 0.91 |
| 30 | 0.75 | 0.72 | 0.83 |
| 35 | 0.63 | 0.59 | 0.73 |
| 40 | 0.52 | 0.47 | 0.65 |
| 45 | 0.45 | 0.40 | 0.58 |
| 50 | 0.39 | 0.35 | 0.53 |
| 55 | 0.36 | 0.32 | 0.49 |
| 60 | 0.33 | 0.30 | 0.47 |
| 65 | 0.31 | 0.28 | 0.44 |
| 70 | 0.29 | 0.27 | 0.43 |
| 75 | 0.28 | 0.26 | 0.41 |
| 80 | 0.27 | 0.26 | 0.40 |
| 85 | 0.26 | 0.25 | 0.39 |
| 90 | 0.26 | 0.25 | 0.39 |
| 95 | 0.25 | 0.24 | 0.38 |

Table 2

Experimental flux-averaged cross-section for the $^{nat}$Mo(γ,xn)$^{93m}$Mo reaction

| $E_{\gamma max}$, MeV | $\langle\sigma(E_{\gamma max})\rangle \pm \Delta\langle\sigma(E_{\gamma max})\rangle$, μb |
|---|---|
| 37.50 | 2.24 ± 0.45 |
| 41.10 | 2.43 ± 0.37 |
| 44.20 | 2.98 ± 0.39 |
| 49.00 | 3.52 ± 0.36 |
| 53.70 | 3.98 ± 0.52 |
| 59.60 | 4.65 ± 0.46 |
| 63.30 | 4.94 ± 0.59 |
| 65.50 | 5.67 ± 0.67 |
| 68.25 | 5.43 ± 0.64 |
| 73.10 | 6.95 ± 0.74 |
| 77.50 | 6.53 ± 0.60 |
| 82.50 | 6.60 ± 0.65 |
| 86.90 | 8.20 ± 0.80 |
| 92.50 | 7.53 ± 0.65 |
| 92.90 | 7.86 ± 0.78 |

## CONCLUSIONS

In the present work, the experiment was performed using the beam of a linear electron accelerator LUE-40 RDC "Accelerator" NSC KIPT and the γ-activation technique. The bremsstrahlung flux-averaged cross-section $\langle\sigma(E_{\gamma max})\rangle_m$ for the photoproduction of $^{93m}$Mo in reactions on natural Mo targets was determined for the first time. The bremsstrahlung end-point energy range was $E_{\gamma max}$ = 38…93 MeV.

The calculation of the flux-averaged cross-sections $\langle\sigma(E_{\gamma max})\rangle_{th}$ was performed using the cross-sections $\sigma(E_\gamma)$ for the studied reactions from the TALYS1.96 code with different gamma strength functions GSF 1–9 and level density models LD 1–6. The calculation result with the parameters LD6 and GSF2 is the closest to the experimental data.

Photonuclear reaction data with the formation of nuclei in a metastable state and the case of reactions with the escape of a large number of particles play an important role in the verification of modern theoretical models. Continuation of the works aimed at measuring the cross-sections of such reactions is very important at present and is possible with the use of high-intensity bremsstrahlung flux of a linear accelerator.

## ПЕРЕРІЗИ ФОТОНЕЙТРОННОЇ РЕАКЦІЇ $^{nat}$Mo(γ,xn)$^{93m}$Mo ПРИ ЕНЕРГІЯХ ГАЛЬМІВНОГО ВИПРОМІНЮВАННЯ ДО 95 МеВ

*І.С. Тімченко, О.С. Деєв, С.М. Олійник, С.М. Потін, В.А. Кушнір, В.В. Митроченко, С.О. Пережогін*

Фотоутворення ядра $^{93m}$Mo на натуральному молібдені досліджено на електронному пучку ЛПЕ-40 НДК «Прискорювач» ННЦ ХФТІ. Вимірювання проводили за допомогою активаційного методу та офлайнової γ-спектрометрії. Експериментальний усереднений за потоком поперечний переріз ⟨σ($E_{γmax}$)⟩$_m$ у діапазоні кінцевих енергій гальмівного випромінювання 38…93 МеВ було вперше визначено для реакції $^{nat}$Mo(γ,xn)$^{93m}$Mo. Теоретичні значення виходів $Y_{m,g}$($E_{γmax}$) та усереднених за потоком перерізів ⟨σ($E_{γmax}$)⟩$_{m,g}$ для реакцій $^{nat}$Mo(γ,xn)$^{93m,g}$Mo були розраховані з використанням перерізів σ($E_γ$) з коду TALYS1.96 для різних гамма-силових функцій та моделей щільності рівнів. Порівняння показало значні розбіжності між експериментальними значеннями ⟨σ($E_{γmax}$)⟩$_m$ і теоретичними результатами.